\documentclass[3p,times,12pt]{elsarticle}

\usepackage{amssymb}

\biboptions{comma,round}

\usepackage[figuresright]{rotating}

\journal{Nuclear Physics A}

\begin{document}

\begin{frontmatter}


\title{Triangular flow of negative pions emitted in PbAu collisions 
at $\sqrt{s_{NN}} = $ 17.3~GeV\tnoteref{t1}}
\tnotetext[t1]{CERES Collaboration}
\author[a]{D.~Adamov{\'a}}
\author[b,k]{G.~Agakichiev}
\author[c,l]{A.~Andronic}
\author[d]{D.~Anto{\'n}czyk}
\author[d]{H.~Appelsh{\"a}user}
\author[b]{V.~Belaga}
\author[e,f,m]{J.~Biel\v{c}\'{\i}kov{\'a}}
\author[c,l]{P.~Braun-Munzinger}
\author[f]{O.~Busch}
\author[g]{A.~Cherlin}
\author[f]{S.~Damjanovi{\'c}}
\author[h,s]{T.~Dietel}
\author[f]{L.~Dietrich}
\author[i]{A.~Drees}
\author[f]{W.~Dubitzky}
\author[f,n]{S.~I.~Esumi}
\author[f,o]{K.~Filimonov}
\author[b]{K.~Fomenko}
\author[g]{Z.~Fraenkel\corref{cor1}}
\author[c]{C.~Garabatos}
\author[f]{P.~Gl{\"a}ssel}
\author[c]{G.~Hering}
\author[c]{J.~Holeczek}
\author[c]{M.~Kalisky}
\author[t,u]{Iu.~Karpenko}
\author[f]{G.~Krobath}
\author[a]{V.~Kushpil}
\author[c]{A.~Maas}
\author[c,l]{A.~Mar\'{\i}n}
\author[f,p]{J.~Milo\v{s}evi{\'c}\corref{cor0}}
\ead{Jovan.Milosevic@cern.ch}
\author[c,l]{D.~Mi{\'s}kowiec}
\author[b]{Y.~Panebrattsev}
\author[b]{O.~Petchenova}
\author[f,q]{V.~Petr{\'a}\v{c}ek}
\author[c]{S.~Radomski}
\author[c,r]{J.~Rak}
\author[g]{I.~Ravinovich}
\author[j]{P.~Rehak\corref{cor1}}
\author[c]{H.~Sako}
\author[f]{W.~Schmitz}
\author[d]{S.~Schuchmann}
\author[c]{S.~Sedykh}
\author[b]{S.~Shimansky}
\author[f]{J.~Stachel}
\author[a]{M.~\v{S}umbera}
\author[f]{H.~Tilsner}
\author[g]{I.~Tserruya}
\author[c]{G.~Tsiledakis}
\author[h]{J.\thinspace P.~Wessels}
\author[f]{T.~Wienold}
\author[e]{J.\thinspace P.~Wurm}
\author[f,c]{S.~Yurevich}
\author[b]{V.~Yurevich}
\cortext[cor1]{\it deceased}
\cortext[cor0]{corresponding author}
\address[a]{Nuclear Physics Institute, Academy of Sciences of the Chech
Republic, 25068 \v{R}e\v{z}, Czech Republic}
\address[b]{Joint Institute of Nuclear Research, Dubna, 141980 Moscow Region,
 Russia}
\address[c]{Institut f\"ur Kernphysik, GSI, 64291 Darmstadt, Germany}
\address[d]{Institut f\"ur Kernphysik, Johann Wolfgang Goethe-Universit{\"a}t 
Frankfurt, 60438 Frankfurt, Germany}
\address[e]{Max-Planck-Institut f{\"u}r Kernphysik, 69117 Heidelberg, Germany} 
\address[f]{Physikalisches Institut, Universit{\"a}t Heidelberg, 
69120 Heidelberg, Germany} 
\address[g]{Department of Particle Physics, Weizmann Institute, Rehovot, 
76100 Israel} 
\address[h]{Institut f\"ur Kernphysik, Universit{\"a}t M{\"u}nster, 48149 
M\"unster, Germany} 
\address[i]{Department for Physics and Astronomy, SUNY Stony
Brook, NY 11974, USA} 
\address[j]{Instrumentation Division, Brookhaven National Laboratory,
Upton, NY 11973-5000, USA}
\address[t]{Frankfurt Institute for Advanced Studies, Ruth-Moufang-Stra{\ss}e
 1, D-60438 Frankfurt am Main, Germany}
\address[k]{Present affiliation:~ II.~Physikalisches Institut der Justus Liebig 
Universit\"at,\\ 35392 Giessen, Germany}
\address[l]{Present affiliation:~ Research Division and Extreme Matter 
Institute (EMMI),\\
GSI Helmholtzzentrum f\"ur Schwerionenforschung, 64291 Darmstadt, Germany}
\address[m]{Present affiliation:~ Nuclear Physics Institute, Academy of 
Sciences of the Czech Republic,\\ 25068 \v{R}e\v{z}, Czech Republic}
\address[n]{Present affiliation:~ Institute of Physics, University of Tsukuba,
 Tsukuba, Japan}
\address[o]{Present affiliation:~ Physics Department, University of 
California,\\ 
Berkeley, CA 94720-7300, USA}
\address[p]{Present affiliation:~ University of Belgrade, Faculty of Physics 
and Vin$\check{c}$a Institute of Nuclear Sciences, \\ 11001 Belgrade, Serbia}
\address[q]{Present affiliation:~ Faculty of Nuclear Science and Engineering,
Czech Technical University,\\ 16636 Prague, Czech Republic}
\address[r]{Present affiliation:~ Department of Physics, University of 
Jyv\"askyl\"a, Jyv\"askyl\"a, Finland}
\address[s]{Present affiliation:~ Department of Physics, 
University of Cape Town, Rondebosch 7701, South Africa}
\address[u]{Present affiliation:~ INFN - Sezione di Firenze, Via G. Sansone 1, I-50019 Sesto Fiorentino (Firenze), Italy} 

\begin{abstract}

Differential triangular flow, $v_3(p_T)$, of negative pions is
measured at $\sqrt{s_{NN}}$= 17.3~GeV around midrapidity by the
CERES/NA45 experiment at CERN in central PbAu collisions in the range 
0-30\% with a mean centrality of 5.5\%. This is the first measurement 
as a function of transverse momentum of the triangular flow at SPS 
energies. The $p_T$ range extends from about 0.05~GeV/c to more than 
2~GeV/c. The triangular flow magnitude, corrected for the HBT effects, 
is smaller by a factor of about 2 than the one measured by the PHENIX 
experiment at RHIC and the ALICE experiment at the LHC. Within the 
analyzed range of central collisions no significant centrality 
dependence is observed. The data are found to be well described by a 
viscous hydrodynamic calculation combined with an UrQMD cascade model 
for the late stages.

\end{abstract}


\end{frontmatter}


\section{Introduction}
\label{introduction}

The azimuthal anisotropy of particles emitted in heavy-ion collisions
is used to study properties of hot and dense systems created in such
collisions. The almond shape of the overlapping region in a
non-central collision manifests itself in the appearance of the
elliptic flow anisotropy~\cite{Ollit92} driven by strong interactions
among constituents of the expanding medium. By these interactions the
geometrical anisotropy of the overlap zone evolves, following the 
pressure gradients, into the momentum
space anisotropy that is measured by the second harmonic coefficient
$v_2$. But due to fluctuating positions of the colliding nucleons, the
event plane derived from the elliptic anisotropy is not a strict plane
of symmetry, and higher-order anisotropies may
appear~\cite{roland}. In fact, among the prominent results from
collider experiments are observations of significant triangular flow,
at the Relativistic Heavy Ion Collider (RHIC) at nucleon-nucleon
center-of-mass energy up to $\sqrt{s_{\rm NN}}=200$~GeV
~\cite{phenix11,star14,star16}, and at the Large Hadron Collider (LHC) at
$\sqrt{s_{\rm NN}}= 2.76$~TeV~\cite{ALICE11,ATLAS12,CMS14}, both in
central and non-central collisions.

The large $v_2$ values of collective flow agree well with predictions
of relativistic hydrodynamics~\cite{HuoRuu06} without
dissipation. This suggests that elliptic flow is developed in the
early phase of a locally equilibrated, strongly
interacting Quark Gluon Plasma (QGP). The QGP behaves as a nearly
perfect liquid with a very small ratio $\eta/s$ of shear viscosity to
entropy density, close to its string-theoretical limit of
$1/4\pi$~\cite{GyuMcL05,HHKLN06}.

 The average elliptic flow magnitude, $v_2$, is about 20\% larger at
 the LHC compared to RHIC~\cite{Aamodt11,Atlas1108,Tserruya11}.  This
 increase is mainly due to the harder $p_{\rm T}$ spectrum at LHC
 energies. The measured $v_2$ is in agreement with hydrodynamical
 extrapolations from RHIC data using the same $\eta/s$
 value~\cite{Luzum11,SHHS11} and also in agreement with a hybrid
 calculation treating the QGP by ideal hydrodynamics and the late
 stages by a hadronic cascade model~\cite{HHN10}. Contrary to the
 elliptic flow, the triangular flow is nearly independent of
 centrality. The triangular flow can be described using viscous 
hydrodynamics and transport models.
 Triangular flow is found to be a sensitive probe of initial geometry
 fluctuations and viscosity~\cite{Alver10}.

The elliptic flow magnitude $v_2$ measured at the Super Proton
Synchrotron (SPS) energy, $\sqrt{s_{\rm NN}}=17.3$~GeV, is about 30\%
lower than those at the top RHIC energy of 
$\sqrt{s_{\rm NN}}=200$~GeV~\cite{phenix11}. 
Except for the most central
collisions~\cite{Adam12}, the differential flow data $v_2(p_{\rm T})$ at
SPS~\cite{NA4903,Aga04,Aggar04}, although very similar in shape to the
RHIC and LHC data, stay below calculations of ideal
hydrodynamics~\cite{Huo_plb01}. This failure of ideal hydrodynamics at
the top SPS energy has been ascribed to insufficient number densities
at very early collision stages~\cite{Heinz04} and strong dissipative
effects at the late hadronic
stages~\cite{GyuMcL05,HHKLN06,Teaney03,Niemi11}.

In this paper, we present the first measurement of triangular flow at
SPS energy. Experimental results comprise differential triangular flow
$v_3(p_{\rm T})$ of negative pions emitted from central 158~$A$GeV
PbAu collisions. The results are compared with the measurement of the
triangular flow performed by the PHENIX collaboration at RHIC and the
ALICE Collaboration at LHC and also with a hydrodynamics calculation coupled
with a UrQMD cascade model~\cite{Karp15} to describe the late
stages. These findings might shed some light on the late stage of
collective expansion characterized by rescattering in the `hadronic
corona'~\cite{HirGyu_NPA06}.

\section{Experiment and data sample}
\label{experiment}

A sample of $30\cdot 10^{6}$ central PbAu events was collected with
the upgraded CERES/NA45 spectrometer during the heavy-ion run at the
top SPS energy of 158 $A$GeV. Within the polar angle acceptance of 
$7.7^\circ < \vartheta <14.7^\circ$, which corresponds to a
pseudorapidity range $2.05< \eta < 2.70$ near midrapidity
($y_{\rm mid}=2.91$), the CERES spectrometer has axial symmetry around
the beam direction.  As it covers the full azimuth $\phi$, it
is very suitable for studies of azimuthal anisotropy. A detailed
description of the CERES experiment is given in \cite{Marin04}.

A precise momentum determination is provided by the radial-drift Time
Projection Chamber (TPC)~\cite{NIM08} which is operated inside an
axially symmetric magnetic field with a radial component providing
deflection in $r_{\phi}$. Negative pions are identified using the
differential energy loss d$E$/d$x$ along their tracks in the TPC. For
vertex reconstruction and tracking outside the magnetic field, two
radial Silicon Drift Detectors (SDD)~\cite{Holl96} are placed at 10
and 13~cm downstream of a segmented Au target. Negative pions are
reconstructed by matching track segments in the SDD doublet and in the
TPC using a momentum-dependent matching window. Depending on pion
momentum, the relative momentum resolution varies between 2\% and 8\%.

\begin{figure}[b!]
 \centerline{\includegraphics[width=14.0cm]
 {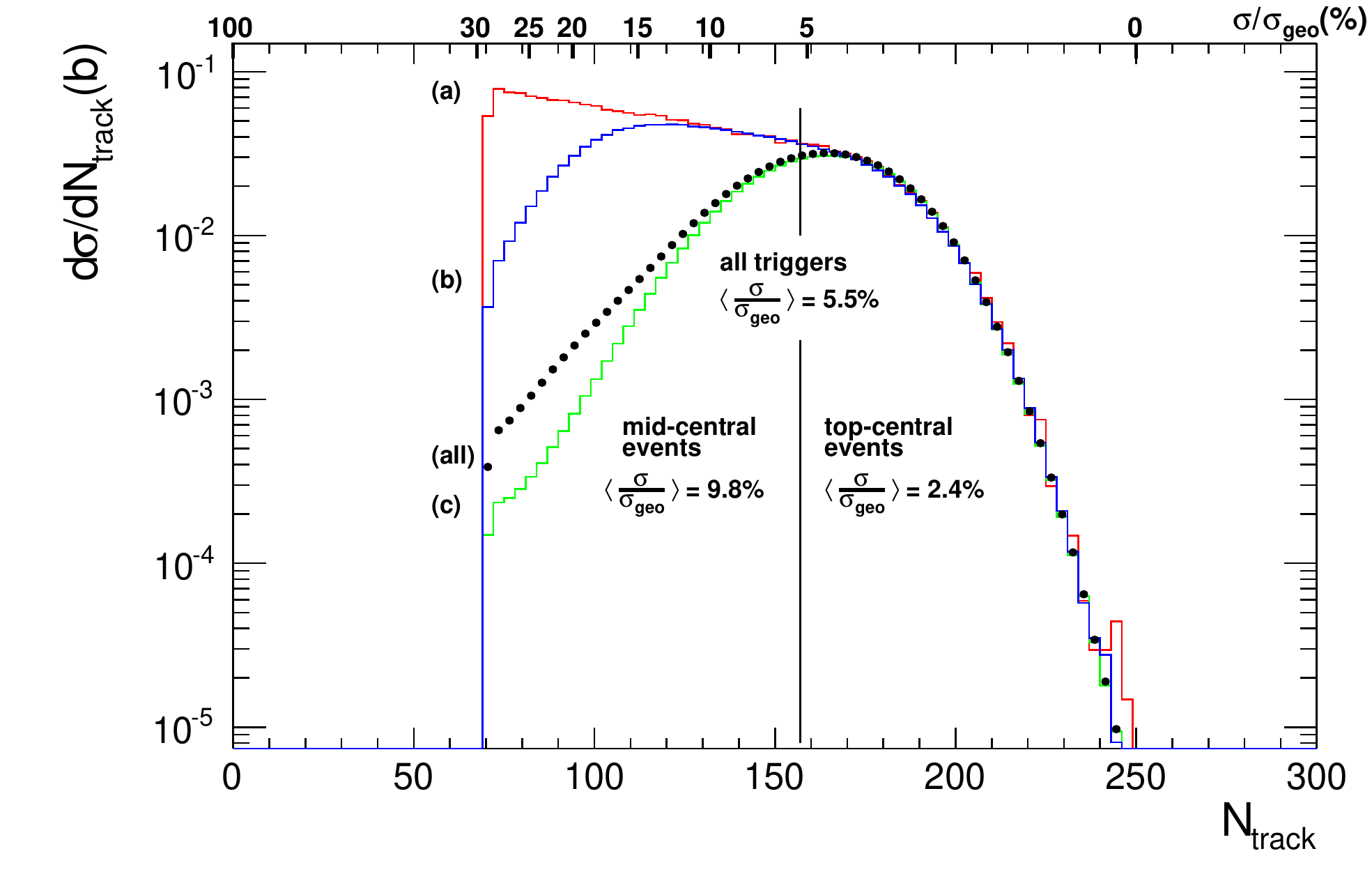}}
  \caption{TPC track density for our trigger mix within (0 -- 30\%)
    centrality. The distribution consists of several components: 
    (a) minimum-bias (0.5\%), (b) semicentral (8.3\%), and 
    (c) central (91.2\%), where the parenthesis represent the 
    percentage fractions in the mix. The mix of all triggers, with 
    a resulting mean centrality of 5.5~\%, is
    labeled `all triggers' and displayed by black circles. 
    The vertical axis represents the differential cross-section 
    expressed in barns (b). The $\langle\sigma/\sigma_{geo}\rangle$ 
    axis on top applies to minimum-bias data only.
 \label{fig:mult_NPA}}
 \end{figure}

A mix of three triggers designed to enhance central events has been
used for data collection in the range \mbox{0 -- 30\%} of
$\sigma/\sigma_{\rm geo}$ with an average centrality of 5.5\% in the
data sample. The track-multiplicity distribution for all triggers
combined (`all triggers') is shown in Fig.~\ref{fig:mult_NPA} by full
black symbols. At low multiplicities, it strongly deviates from the
minimum-bias distribution labeled $(a)$. Beside minimum-bias data,
which contribute 0.5\%, a semi-central trigger, labeled $(b)$,
contributes 8.3~\% to the total. The biggest share of data, 91.2\%, is
collected with a most central trigger labeled $(c)$ in
Fig.~\ref{fig:mult_NPA}. Note that the data will be presented here,
besides `all triggers', for `top-central' and `mid-central' triggers
by selecting $N_{track}> 159$ or $\leq 159$, with weighted mean
centralities of 2.4\% and 9.8\%, respectively.  We remark that
because of the unconventional shape of the `all triggers' and
`mid-central' distributions, we have supplied the actual distributions 
in digitized form for theory comparisons.

\section{Analysis and results}
\label{results}

Among the higher-order harmonics, the triangular collective flow is of
particular interest. It is quantified by $v_3$, the third-order
harmonic coefficient of the azimuthal particle distribution measured
with respect to $\Psi_{3}$, the azimuthal angle of the 3rd-order
participant event-plane. The angle $\Psi_{3}$ is determined as:
\begin{equation}
\label{eq:Phi3}
\Psi_{3}=\frac{1}{3}\arctan\frac
{\displaystyle\sum\limits_{i=0}^nw_{i}(p_{Ti})
\sin(3\phi_i)}{\displaystyle\sum\limits_{i=0}^nw_{i}(p_{Ti})\cos(3\phi_i)}.
\end {equation} 
Here, $\phi_i$ is the azimuthal angle of the $i$-th particle out of
$n$ used for event-plane reconstruction, and $w_{i}(p_{Ti})$ are 
weights used to optimize the event-plane resolution. In the same way 
as in~\cite{Adam12}, the $\phi$ coordinates are divided into 100
adjacent equal slices spanning the full azimuth.
 \begin{figure}[h]
\centerline{\includegraphics[width=12cm]{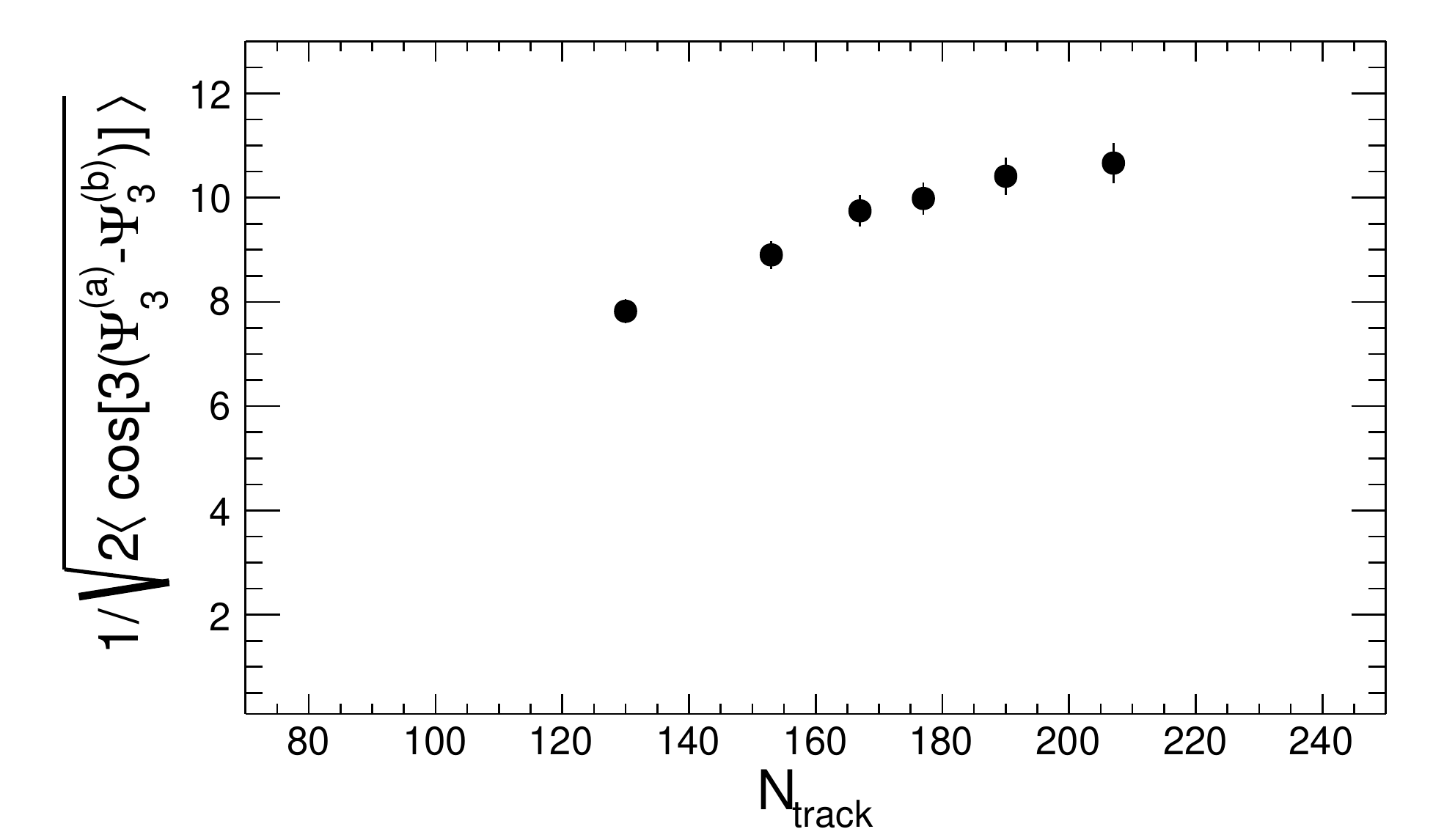}}
  \caption{The correction factor of the 3rd-order event plane as 
a function of TPC multiplicity for the 2-subevents method (pion data). 
  \label{fig:res}}
  \end{figure}
This approach is used to avoid the trivial autocorrelation effect, and
some contribution from short-range correlations.  In order to correct
for local detector inefficiency, a shifting and flattening procedure
has been applied (for more details see~\cite{Jovan06}) to ensure an
azimuthally isotropic event-plane distribution.  The azimuthal
anisotropy of particle tracks is then measured with respect to the
3rd-order event plane reconstructed by employing tracks from
non-adjacent slices only.  The finite resolution of the event plane
orientation is obtained from the differences between the event planes
reconstructed from two sliced subevents, $a$ and $b$. The
corresponding correction factor is calculated as $(2\langle
\cos[3(\Psi^{(a)}_{3}-\Psi^{(b)}_{3})]\rangle)^{-1/2}$, and used to
compensate the raw $v_3$ for finite event-plane resolution.  As the
latter depends on multiplicity, the correction factor is calculated
for different centralities. Both, the event multiplicity and the
$v_{3}$ magnitude influence the event-plane resolution.
Fig.~\ref{fig:res} shows that a decrease of the dispersion in
event-plane orientation with increasing multiplicity is weaker than
the decrease in anisotropy.
In order to reduce statistical errors, the $v_{3}$
results, presented in this paper, are obtained by merging the results
obtained in six narrower multiplicity bins.

Since almost all particles accepted for analysis are negative pions,
subsamples become partially correlated due to the Hanbury Brown \& Twiss
effect (HBT) of identical bosons. This effect produces a
space-momentum correlation between two pions of the same charge if the
product of their momentum difference and the source radius $R$ is
below the uncertainty limit, i.e., $|\vec p_2 - \vec p_1|\leq
\hbar/R$. In the rather central collisions under study here, $R$ is
typically 7~fm, and consequently $\hbar/R\approx 30$~MeV/c, much
smaller than the mean pion momentum $\langle p_{\rm T}\rangle\approx
400$~MeV/c. Moreover, the HBT correlation is short range also in
azimuth, and it is significant only if $|\phi_1-\phi_2|\le \hbar/Rp_{\rm
  T}\simeq~0.1$. As we deal with bosons, the correlation is positive
like flow itself, and therefore applying the flow analysis to the HBT
correlations would result in a spurious flow.

In order to subtract the non-flow HBT contribution we follow
~\cite{Dinh00,Jana03} and use the standard Bertsch-Pratt
parametrization in the comoving system. The corresponding parameters
$R_{side}$, $R_{out}$, $R_{long}$ describe the dimensions of the
source and the so-called chaoticity parameter $\lambda$ their degree
of non-coherence. $\lambda$ is allowed to be varied by generous
$\pm$50\% to account for different track resolution in TPC and the SDD
doublet; the former is used for the determination of the source
parameters, the latter for the event plane determination.
The numerical values $R_{side}$, $R_{out}$, $R_{long}$ and $\lambda$ are 
obtained from the CERES HBT data~\cite{Tilsner02,adamova03,HBTCERES} by 
averaging over $k_{t}\leq$~0.6~GeV/c. 

\begin{figure}[t!]
\centerline{\includegraphics[width=12cm]{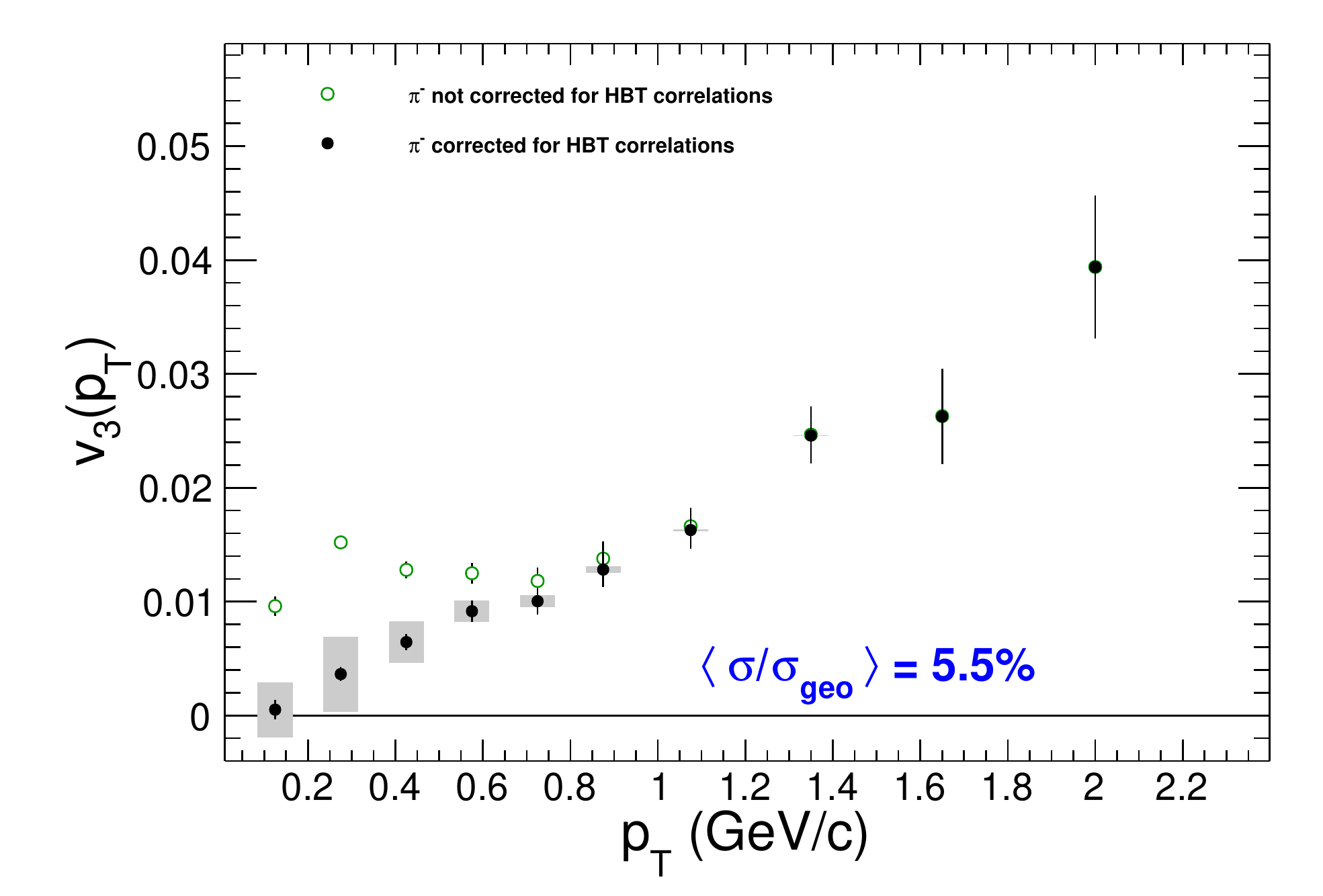}}
  \caption{The magnitude $v_3$ of triangular flow as a function of 
   negative pion transverse momentum before (open green circles) and 
   after (closed circles) correcting for the HBT effect. `All 
   triggers', averaged centrality 5.5\%. Statistical uncertainties are
   represented with the error bars, while systematic ones are indicated 
   by gray rectangles.
  \label{fig:hbt}}
  \end{figure}
 
As expected, the size of the corrections applied to the data displayed
in Fig.~\ref{fig:hbt} is quite large at low $p_{\rm T}$, but decreases
rapidly with increasing $p_{\rm T}$. The systematic uncertainty in the HBT
contribution is derived by calculating the correction varying all
source parameters by $\pm 1~\sigma$ together and independently,
and then taking the error of the mean of the resulting distribution to
represent the systematic uncertainty in each bin. 

The systematic uncertainties in the corrected $v_3$ have significant size (up
to 0.4\%) just in the $p_{T}$ region where the HBT effect is
greatest. They become negligible for $p_{T}$ approaching 0.8~GeV/c.
In order to stabilize the final HBT-corrected value of the triangular
flow, several iterations of the correction procedure described 
in~\cite{Dinh00} have been performed until the difference between the final
HBT-corrected $v_{3}$ value and the one before it became smaller than 
$10^{-4}$. The triangular flow values corrected this way increase about 
linearly, starting from zero at transverse momenta close to zero up to 0.04 
at $p_{T}$ around 2~GeV/c.

\begin{figure}[t!]
  \centerline{\includegraphics[width=12cm]{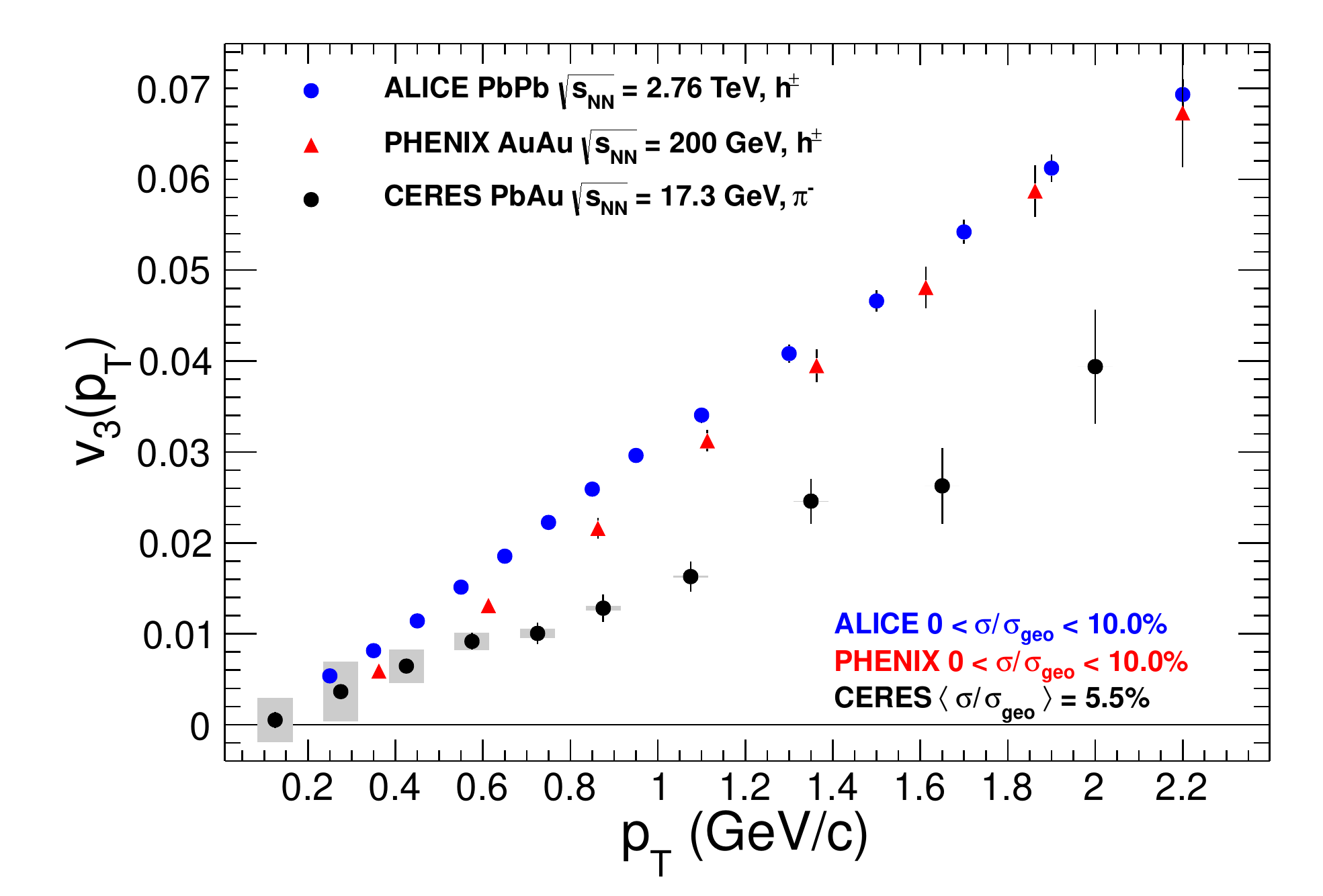}}
  \caption{Comparison of triangular flow $v_{3}$ of negative pions
    $vs$ $p_{T}$: from PbAu collisions at $\sqrt{s_{\rm NN}}=17.3$~GeV
    (CERES, solid black circles); from AuAu collisions at
    $\sqrt{s_{\rm NN}}=200$~GeV (PHENIX, red triangles); from PbPb
    collisions at $\sqrt{s_{\rm NN}}=2.76$~TeV (ALICE, solid blue circles)
    at comparable centrality. Statistical uncertainties are
   represented with the error bars, while systematic uncertainties of the
   CERES results are indicated by gray rectangles.
\label{fig:comp}}
\end{figure}

Fig.~\ref{fig:comp} compares our triangular flow results with those
from the PHENIX and ALICE Collaborations at $\sqrt{s_{\rm NN}}=200$~GeV 
and $\sqrt{s_{\rm NN}}=2.76$~TeV~\cite{phenix11,ALICEtriang},
respectively, in the limited $p_{\rm T}$ range accessible to CERES and
at comparable centrality.

By inspection of Fig.~\ref{fig:comp} we conclude that the magnitudes
of triangular flow at RHIC and at LHC energy are nearly
equal~\cite{Gale}. In contrast, the magnitude at the top SPS energy 
reaches only about one half of the corresponding value at LHC
energy. The transverse momentum range of the analyzed SPS data is
small with respect to that covered by ALICE
data~\cite{ALICEtriang}. In this restricted $p_{T}$ range the data suggest
a linear $v_3(p_{\rm T})$ dependence starting from zero.

We like to remark that ALICE uses large gaps in
pseudo-rapidity between tracks used for event-plane reconstruction and
tracks to measure $v_3$; this way non-flow contributions from jets and
mini-jets might have been effectively suppressed. Although jet-like
correlations have been observed at SPS energy \cite{Aga04} at the much
lower $\sqrt{s_{\rm NN}}$ compared to LHC, the minijet density is strongly
reduced. In Ref~\cite{Aga04} it is shown that in very central PbAu collisions 
at SPS energies the total jet yield is about 0.02 per event which is more 
than an order of magnitude 
smaller with respect to the corresponding yield at the LHC energy, while the
charged particle pseudo-rapidity density is only 4 times 
smaller~\cite{alice2010}. This is 
quite fortunate, since to employ a pseudo-rapidity gap 
is no option for the limited acceptance in CERES. 

\begin{figure}[t!]
  \centerline{\includegraphics[width=12cm]{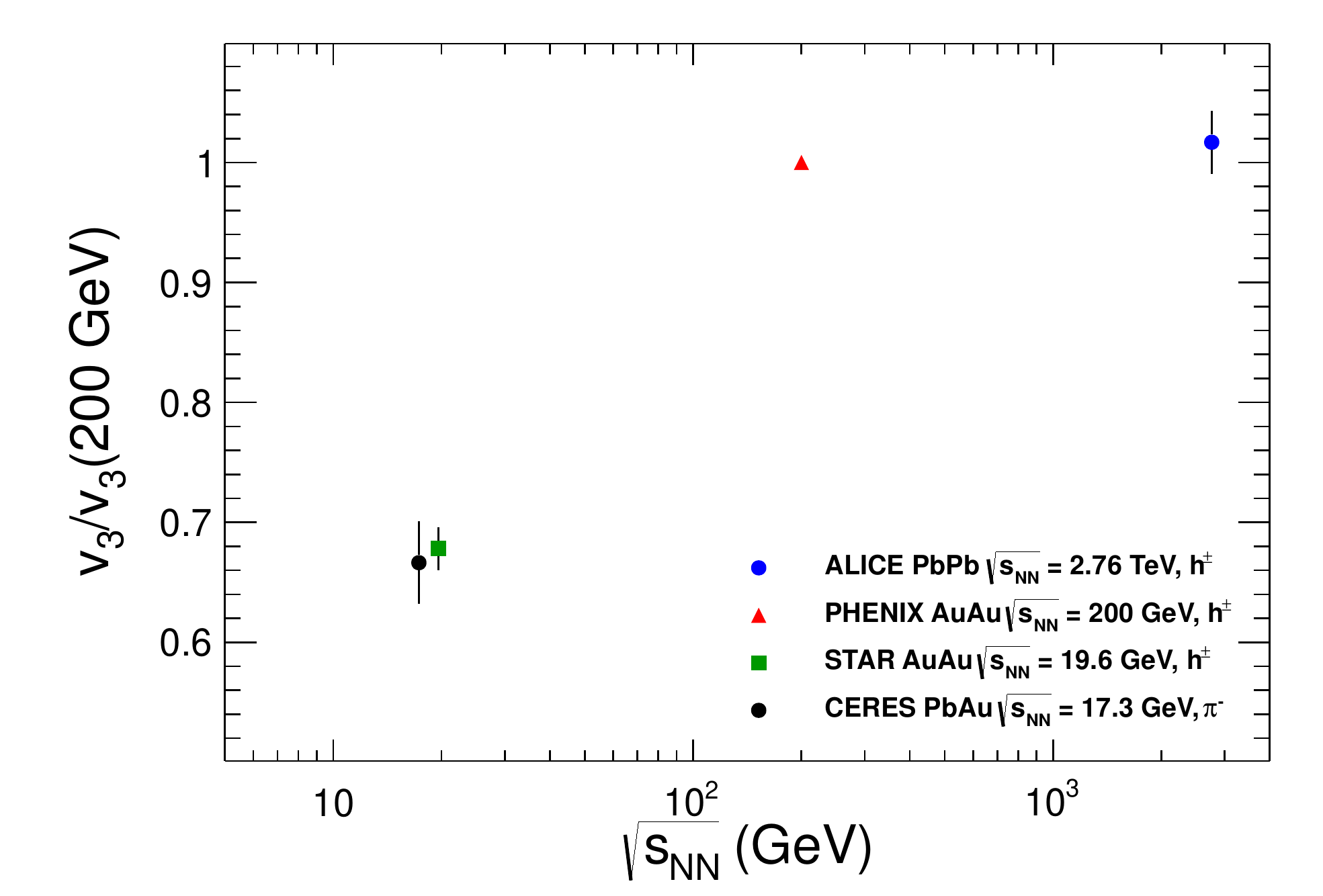}}
  \caption{Ratios of the triangular flow $v_{3}$ measured at different 
collision energies with respect to the $v_{3}$ magnitude measured at PHENIX AuAu collisions at $\sqrt{s_{\rm NN}}=200$~GeV. The $v_{3}$ are obtained by integrating the corresponding differential $v_{3}(p_{T})$ over 0.3 $< p_{T} <$ 2.1~GeV/c. The data from ALICE, PHENIX and STAR were taken from~\cite{star16}, ~\cite{phenix11} and ~\cite{ALICEtriang} respectively.
\label{fig:rat}}
\end{figure} 

On Fig.~3 in~\cite{star16} is shown of the $p_{T}$-integrated two-particle Fourier coefficients, i.e. of the squared $v_{3}$ magnitude as a function of $\sqrt{s_{NN}}$ energy with a shallow minimum between 10 and 20~GeV. The integration has been performed for $p_{T} >$~0.2~GeV/c. The ratio between the $v_{3}$ at 19.6~GeV, which is quite close to the top SPS energy of 17.3~GeV/c, with respect to the $v_{3}$ measured at 200~GeV is about 0.63. In Fig.~\ref{fig:rat} are depicted corresponding ratios for 17.3, 19.6, 200 and 2760~GeV/c where the $p_{T}$ integration has been done within the range 0.3 $< p_{T} <$ 2.1~GeV/c. The $v_{3}$ ratio between the top SPS and the top RHIC energy is about 0.66 which is quite close to the one found in~\cite{star16}. This is alo in a rather good agreement with an AMPT predictions from~\cite{ampt} for the ratio of about 0.6. 
  
In contrast to elliptic flow which reflects the initial anisotropy of
the fireball and thus depends strongly on centrality (see Fig.~24
in~\cite{Adam12} and Fig.~6.22 in \cite{Jovan06}), triangular flow
arises entirely from fluctuations of the initial shape, and we see
\begin{figure}[t!]
  \centerline{\includegraphics[width=12cm]
{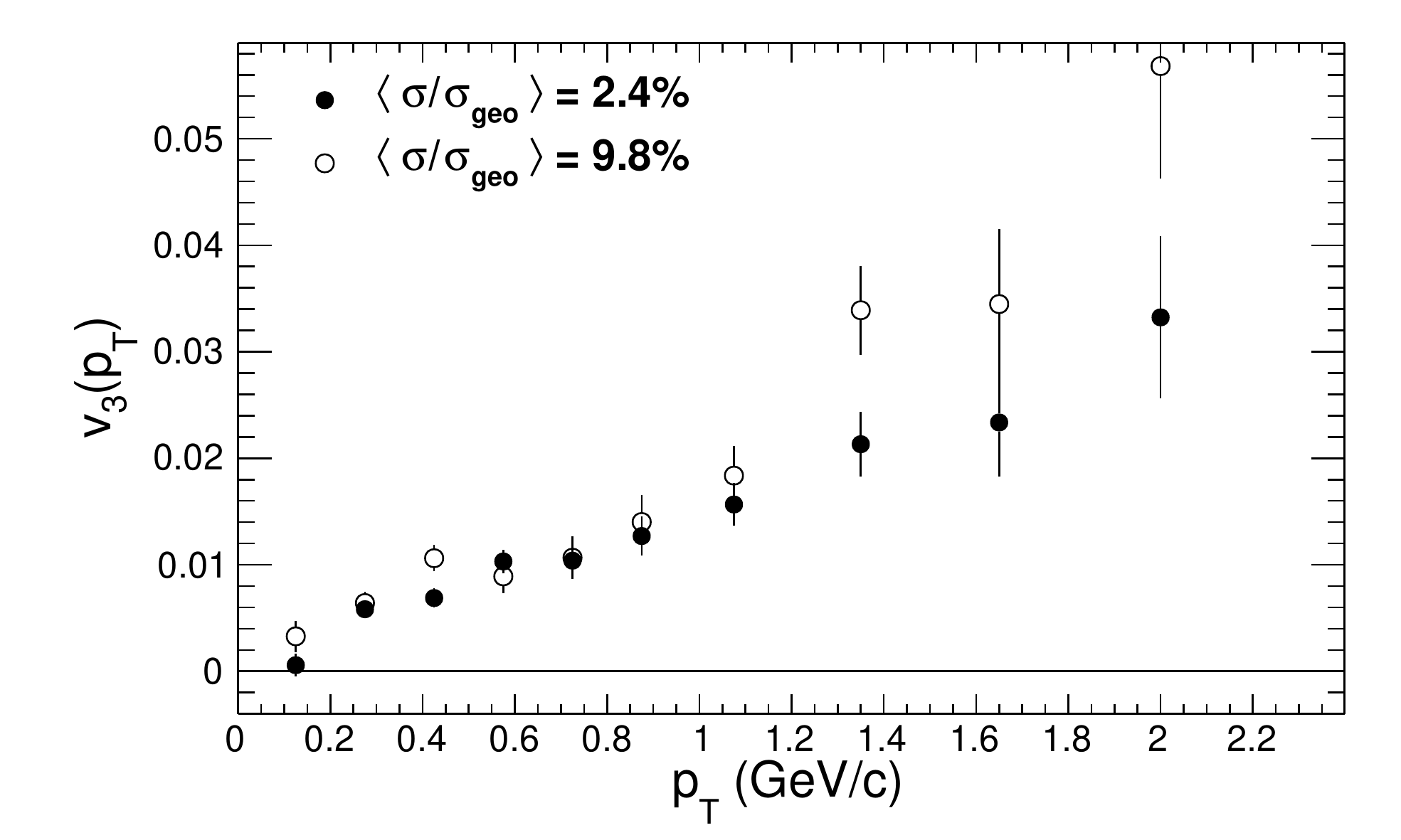}}
  \caption{Triangular flow $v_{3}(p_{\rm T})$ at two different mean
    centralities, at top-central (closed circles) and at mid-central
    (open circles) collisions.
\label{fig:cent}}
  \end{figure} 
\noindent
from Fig.~\ref{fig:cent} that its magnitude is not
significantly different for mid-central and top-central collisions,
with mean averaged  centralities of 2.4\% and 9.8\%, respectively.
A rather weak centrality dependence has also been reported by ALICE
(see Fig.~1 in \cite{ALICEtriang,Gale}) where a
very slight increase of $v_{3}$ with centrality has been observed.
\begin{figure}[t!]
  \centerline{\includegraphics[width=14.5cm]
{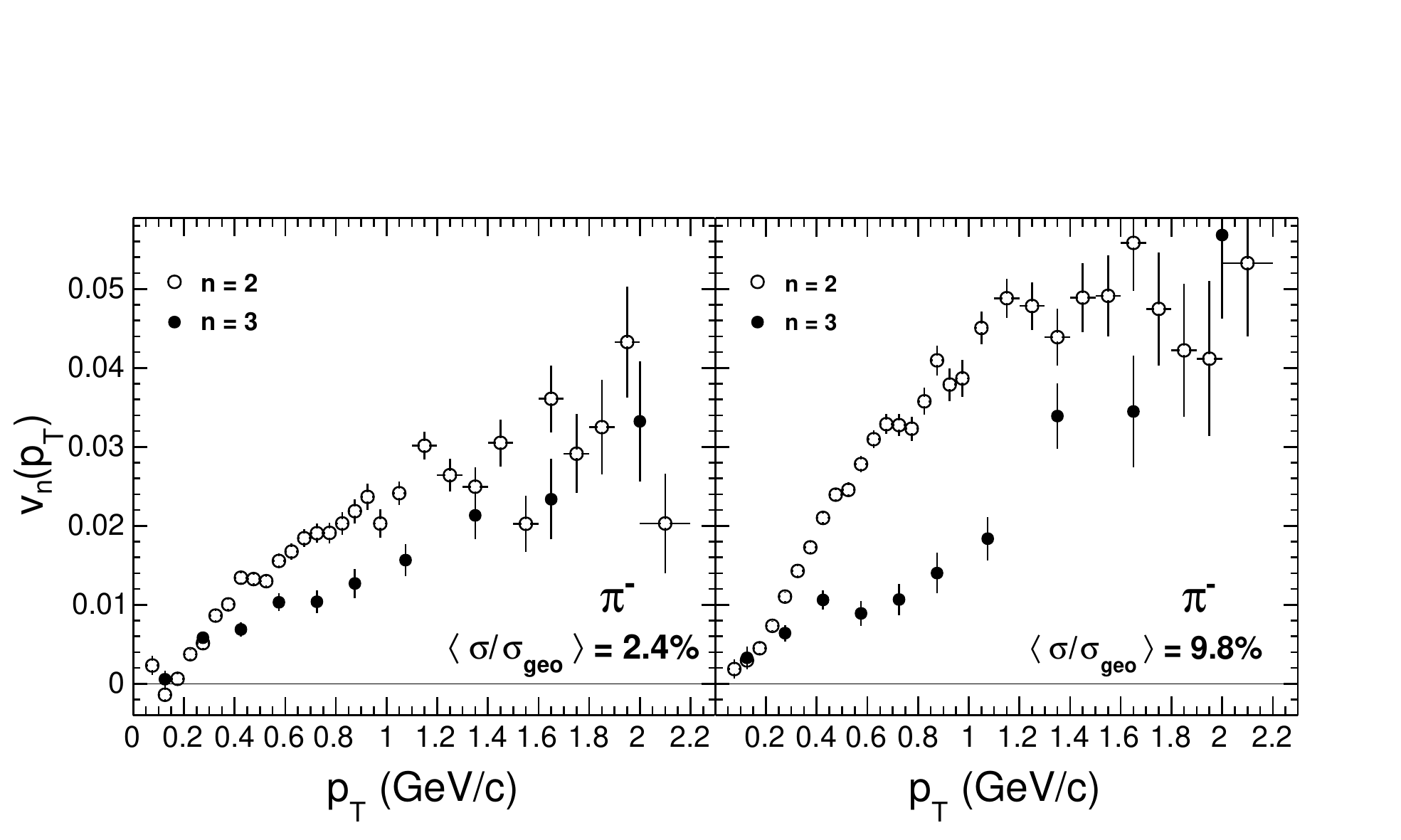}}
  \caption{Elliptic and triangular flow magnitudes, $v_2(p_{\rm T})$ (open
    circles) and $v_3(p_{\rm T})$ (closed circles), respectively, for 
    top-central (left panel) and mid-central collisions (right panel).  
\label{fig:twocent}}
  \end{figure} 
\noindent
The different centrality behaviour of elliptic and triangular flow 
can also be observed from the corresponding $p_{\rm T}$-dependencies displayed 
in  Fig.~\ref{fig:twocent}. The systematic errors of the $v_{3}$ data shown 
in Fig.~\ref{fig:cent} and Fig.~\ref{fig:twocent} are 
very similar to those found for mean centrality of 5.5\%.


Approaches combining the relativistic hydrodynamics with transport
models (so-called hybrid models) have been applied to describe the
expansion stage of heavy-ion collisions at ultra-relativistic
energies.  In such models viscous or ideal fluid dynamics is used to
describe the evolution of the hot and dense quark-gluon plasma, 
and hadron transport to describe the evolution of the
late sparse hadron gas. The calculations shown here are done using a
hybrid model~\cite{Karp15} combining the vHLLE viscous
hydrosolver~\cite{Karpenko:2013wva} with UrQMD hadron
cascade~\cite{Bass:1998ca}. In this model both kinetic and chemical
freeze-outs are described dynamically by the UrQMD hadron cascade, and
thus there are no clear freeze-out temperatures. With this approach, 
particle yields, in particular for strange mesons and baryons, are not well 
described. However, since we deal here with pions only, this may not be a 
serious shortcoming. The switch from
fluid to cascade, the `particlization'~\cite{Karp15}, is set to take
place on a constant energy density surface where $\epsilon=
0.5$~GeV/fm$^{3}$. Since the net baryon density is not uniform on such a
surface, this density does not correspond to a single temperature.
Within the chiral model of the Equation of State (EoS) used in this 
hydrodynamics description, the value of the switching density 
$\epsilon_{sw}$ corresponds to $T \approx $175~MeV at $\mu_B = $0. The 
remaining parameters of the model are the two Gaussian radii for the 
initial distribution of energy, and the starting time for the hydrodynamic 
phase. Their values, together with the
value of the switching density, $\epsilon_{sw} = 0.5$~GeV/fm$^{3}$, are
based on reproduction of the data in collisions at RHIC energies, and
are kept unchanged at the SPS for simplicity. In Ref.~\cite{Karp15}, the 
authors investigated parameter dependence of the model results in the case of 
0-5\% centrality AuAu collisions at $\sqrt{s_{NN}} = $ 19.6~GeV. It is 
shown that the model results change less than 10\% when the parameters 
of the model are varied by 10\%. 
Within the model~\cite{Karp15} is studied the dependence 
of the elliptic and triangular flow magnitude on the collision energy. 

\begin{figure}[h]
  \centerline{\includegraphics[width=12cm]{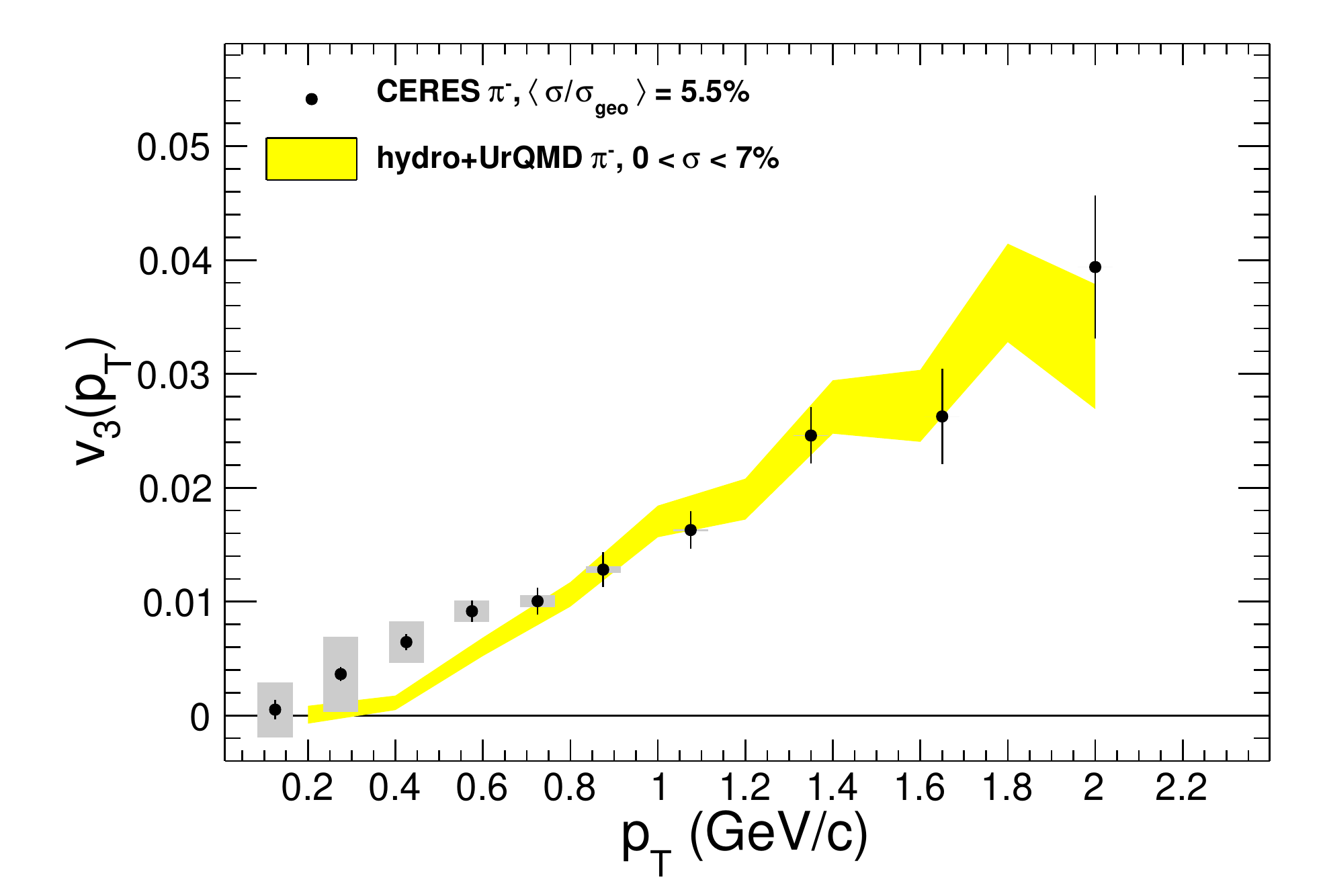}}
  \caption{Comparison between the $p_{T}$ dependence of negative pion
    $v_{3}$ measured in PbAu collisions at $\sqrt{s_{\rm NN}}=17.3$~GeV
    with hydrosolver+UrQMD model predictions. The statistical errors of the
    model predictions are shown as yellow band. Statistical uncertainties 
    of the experimental results are represented with the error bars, while 
    systematic ones are indicated by gray rectangles.
\label{fig:hydro}}
  \end{figure} 

In Fig.~\ref{fig:hydro}, the comparison between the
predictions by this hydrosolver+UrQMD model and our $v_{3}(p_{T})$
measurements of negative pions in PbAu collisions is shown. The model
predictions are calculated for hadrons within $0.2 < p_{T} <
2.0$~GeV/c and  $-1 < \eta < 1$, which is very close
to the experimental acceptance.  Also, the centrality samples which
roughly correspond to the experimental ones are simulated. Comparing
the presented distributions, one can conclude that the model
predictions are in a rather good agreement with the experimental
results, except in the $p_{T}$ region between 0.3 and 0.7~GeV/c where 
the model slightly underpredicts the experimental data.
\section{Summary}
\label{summary}

The triangular flow appears as a hydrodynamic response of the system
created in heavy-ion collision to the fluctuation of the positions of
the overlapping nucleons at the moment of impact. In this paper, for the
first time, results on the differential triangular flow $v_{3}(p_{T})$
are presented measured at the top SPS energy. The magnitudes of $v_3$
are found to be about one half of the ones measured at the top RHIC and LHC
energies. The $v_{3}$ measured by CERES at SPS energy of  
$\sqrt{s_{\rm NN}}=17.3$~GeV is similar to the one measured by STAR at RHIC 
energy of $\sqrt{s_{\rm NN}}=19.6$~GeV. The hydrosolver+UrQMD model is able to 
reproduce the experimental
data rather well. This comparison could shed some light on the dynamics of
the system created in heavy-ion collisions at top SPS energy.

\section{Acknowledgments}

We are grateful to Pasi Huovinen for his guidance concerning the 
hydrodynamical calculations to be compared to SPS data and 
appreciate his critical reading of the manuscript concerning the 
applied hydrosolver+UrQMD model. We acknowledge the support by the 
Ministry of Education, Science and Technological Development of the 
Republic of Serbia throughout the project 171019.





\bibliographystyle{model1-num-names}




\end{document}